\documentclass[printer]{aa}
\usepackage{graphicx}
\usepackage{rotating}
\def\gtrsim{\mathrel{\hbox{\rlap{\hbox{\lower4pt\hbox{$\sim$}}}\hbox{$>$}}}}
\def\ltsim{\mathrel{\hbox{\rlap{\hbox{\lower4pt\hbox{$\sim$}}}\hbox{$<$}}}}

\begin{document}

\title{The magnetic Bp star 36 Lyncis\\II. A spectroscopic analysis of 
its co-rotating disk}
\authorrunning {Smith, Wade, Bohlender, \& Bolton }
\titlerunning{The disk of 36 Lyncis }

\author{M.A. Smith\inst{1}, G.A. Wade\inst{2}, D.A. Bohlender\inst{3}, C.T. Bolton\inst{4}}
\institute{Department of Physics, Catholic University of America, 
Washington, DC 20064, USA; Present address: Space Telescope Science
Institute, 3700 San Martin Dr., Baltimore, MD 21218 ~~Email:~ msmith@stsci.edu \and
Department of Physics, Royal Military College of Canada, Kingston, 
Ontario, Canada K7K 4B4 
\and
National Research Council of Canada, Herzberg Institute of Astrophysics, 5071 W. Saanich Rd., Victoria, BC Canada V9E 2E7 
\and
David Dunlap Observatory, University of Toronto, P.O. Box 360, Richmond Hill, ON, Canada, L4C 4Y6 }

   \date{Received ??; accepted ??}

\abstract{We report on the physical properties of the disk-like structure 
of B8\,IIIp star 36\,Lyncis from line syntheses of phase-resolved, 
high resolution spectra obtained from the 
{\it International Ultraviolet Explorer} archives and from
newly obtained ground-based spectra of the H$\alpha$ absorption profile. 
This disk is highly inclined to the rotational axis and betrays its existence
every half rotation cycle as one of  two opposing sectors pass in front of the 
star. Although the disk absorption spectrum is at least ten times too weak to 
be visible in optical iron lines during these occultations, its properties can 
be readily examined in a large  number of UV ``iron curtain" lines because of
their higher opacities. The UV Fe\,II and Fe\,III lines in particular permit a
determination of the disk temperature: 7,500$\pm{500}$\,K and a column density 
of 3$_{-1.5}$$\hspace*{-.25in}^{+3}$$\hspace*{.1in}\times$10$^{20}$\,cm$^{-2}$.
The analysis of the variations of the UV resonance lines brings out
some interesting details about the radiative properties of the disks: 
(1) they are optically thick in the C\,IV and Si\,IV doublets, 
(2) the range of excitation of the UV resonance lines is larger at the 
primary occultation ($\phi$ = 0.00) than at the secondary one,
and (3) the {\bf relative strengths of the absorption peaks} 
for the two occultations 
varies substantially from line to line. We have modeled the absorptions of 
the UV C\,IV resonance and H$\alpha$ absorptions 
by means of a simulated disk 
with opaque and translucent components. Our simulations suggest that
a gap separates the star and the inner edge of the disk. The disk 
extends radially out to $\geq$10 R$_{*}$. The disk scale height 
perpendicular to the plane is $\approx$1R$_{*}$. 
However, the sector causing the primary occultation is about 
four times thicker than the opposite sector.  The C\,IV scattering region 
extends to a larger height than the H$\alpha$ region does, probably because
it results from shock heating far from the cooler disk plane.
}
\maketitle

%\newpage  
\section{Introduction}
 
  In a companion to this paper, Wade et al. (2006; hereafter Paper 1) 
have redetermined the magnetic and rotational periods 
and surface abundances of the He-weak B8p star 36\,Lyncis (HD\,79158). 
Using data from the {\it International Ultraviolet Explorer (IUE)}
satellite, Shore \& Brown (1990; hereafter SB90) discovered the presence 
of circumstellar plasma through the periodic modulations of the strengths 
of its C\,IV and Si\,IV resonance lines. Following the model of Shore (1987), 
SB90 found that these modulations are due to a disk-like structure 
surrounding the star.\footnote {The term
 ``disk" will be used in this paper to describe circumstellar matter 
 corotating in or near the magnetic equatorial plane of the star. Evidence 
 is accumulating that the gas density of these structures is unevenly 
 distributed in azimuth in this plane. The structure may have 
 a discrete inner edge, may be decentered, may be nonisothermal around the 
 disk {\bf at a given radius from the star's center}, and may be warped out of 
 coplanarity (see e.g., Townsend \& Owocki 2005; hereafter TO05).} 
According to this picture, the wind from the magnetic poles flows out
along open field lines to infinity, much as for unmagnetized stars.
However, at lower magnetic latitudes the wind behavior and geometry is 
complex. Since the wind particles from this zone are charged, 
they cannot easily cross the magnetic loops, and the high velocity flow is 
effectively quenched. At intermediate latitudes, the flow is constrained 
to follow closed magnetic loops. As they cross the magnetic 
equator, the particles collide and shock with streams originating 
from the other hemisphere, generating EUV or X-ray emission in a large, 
spheroidal post-shock region (e.g., Babel 1997, Babel \& Montmerle 1997
(hereafter BM97), Babel \& North 1997). Post-shock 
particles cool and settle toward the equatorial 
plane. Steady state is achieved as particles return to the star's surface
or escape through the disk's outer edge. In this picture a 
cool disk co-rotates with the star and resembles a wobbling inner tube. 
As fixed external observers, we see ``shell" absorption components in the UV 
resonance and H$\alpha$ lines near the rest frame
at those times during the rotation cycle when 
a section of the disk {\bf transits} the star. In spectra of hot 
Bp stars a slightly redshifted emission is observed in the UV lines 
when the disk is viewed ``face on."

  Since the pioneering studies of Babel and collaborators, 
several important advances 
have been made in the hydrodynamic modeling of Bp disks. The first of these 
was the inclusion of the feedback effects of the wind particle weight on 
the magnetic field loops (ud-Doula \& Owocki 2002, uO02). This weight causes
the loops to sag, thereby permitting disk particles to
fall back to the star at low magnetic latitudes. This process leaves an
effectively gas free region close to the star.
In addition, for sufficiently high ratios of magnetic to wind energy 
density $\eta$, circulation of wind efflux proceeds on a timescale 
of a few days and, in certain $\eta$ regimes, chaotically. These models 
suggest that the effective coverage factor of the star by the disk can be 
large, even occulting the entire star for near edge-on viewing 
angles.  Models that include 
radiative cooling of the wind/disk shock interface (Tonneson et al. 2002) 
indicate that the hot zone is mainly confined to the outer region of the disk 
yet also extends out of the equatorial plane. Recent theoretical work 
(e.g., Preuss et al. 2004, TO05) suggests that matter can 
accumulate on low potential energy surfaces, well outside the disk plane.

   A large number of studies have shown that many and perhaps all Bp stars 
with measurable dipolar fields harbor co-rotating magnetospheres. 
Periodic ultraviolet variations have also been found 
in a few $\beta$\,Cep stars (Henrichs et al. 1998, Floquet et al. 2002, 
Neiner et al. 2003), in the O7\,V 
star $\theta^1$\,Ori\,C  (Walborn \& Nichols 1994, Stahl et al. 1996),
and in He-weak stars as late as type B8 (Shore \& Brown
1990b, Shore et al. 2004). To generalize, the
periodic variability of the C\,IV and/or N\,V lines is a hallmark of Bp 
stars with oblique dipolar magnetic fields (Smith \& Groote, hereafter SG01).

  36\,Lyn is one such star. It lies near the cool edge of the He-weak Bp 
domain and shows prominent C\,IV and N\,IV line variations. In order to 
take advantage of the available H$\alpha$ observations and {\it IUE} archival
data, the improved determination of its rotational
and magnetic geometry, and of our improved understanding of the theoretical 
properties of disks, we have modeled the variations of several UV and 
optical lines in this star's disk. From our results we are able to provide 
quantitative estimates of its thermodynamic and geometrical properties. 

  In their analysis of disk properties of four Bp and $\beta$\,Cep stars,
SG01 utilized archival spectra from the {\it International Ultraviolet
Explorer} satellite to analyze the time variations of UV ultraviolet
resonance lines and the 
``iron curtain" background of weak lines populating the far-UV spectrum. 
For this study we were able to include new phased H$\alpha$ 
absorptions, as taken from Paper 1; see the journal of observations in
this paper for details. The finer 
{\bf sampling} in phase and the higher quality of these observations over 
the {\it IUE} data provide additional constraints on the properties of 
the circumstellar material.

  The UV data for these programs are extant high-dispersion {\it IUE}
echellograms obtained through the large-aperture and of the Short Wavelength 
Prime (SWP) camera. These data were obtained from the MAST 
archives.\footnote{Multi-Mission Archive at Space Telescope
Science Institute, in contract to NASA.} and constitute a total of 26
spectra (a 27th spectrum, SWP\,32982, was lost to the {\it IUE} project;
Levay 2003). 

% Table 1 Fundamental parameters
% 
\begin{table}[ht!]
\begin{center}{}

\begin{tabular}{r|l}

\hline

\noalign{\smallskip} 

%Mass  $M$& $4.0\pm 0.2 {\cal M}_\odot$\\
Mass  $M$& $4.0\pm 0.2 M_\odot$\\

%Radius $R$& $3.4\pm 0.7 {\cal R}_\odot$\\
Radius $R$& $3.4\pm 0.7 R_\odot$\\

Effective temperature $T_{\rm eff}$          & $13,300\pm 300$ K\\

Surface gravity $\log g$                  & $3.7-4.2$\\

Period $P_{\rm rot}$      & 3.83475 days\\

Projected rotation speed $v_{\rm e}\sin i$     & $48\pm 3$ km\,s$^{-1}$ \\ 
%True rotation speed $v_e$                             & $45\pm 9$~\kms\\

Inclination $i$           & $56-90^{o}$ \\

Dipole strength $B_{\rm d}$        & 3210-3930 G\\

Phase of mag. equator passage & 0.0 \\

Photospheric abundance [Fe]   & +0.8-1.1 dex \\

\hline\hline\noalign{\smallskip}

\end{tabular}
 \caption{\label{tab1}{Some characteristics of 36 Lyn from Paper 1.} }

\end{center}

%\label{tab:bz}

\end{table}

\section{Spectral synthesis computations}

  To perform the analysis of the {\it IUE} spectra, we utilized a suite  
of programs written by I. Hubeny and collaborators and LTE solar-abundance 
atmospheres computed by Kurucz (1993) for input to these programs. The first 
of these is {\it SYNSPEC,} which is a photospheric line synthesis program 
written by Hubeny, Lanz, \& Jeffery (1994).
For the line synthesis computations using this program, we adopted the
metallic abundances tabulated in Paper\,1. We also employed the features 
in {\it SYNSPEC} that permit the convolution of instrumental 
and rotational broadenings to the computation of the line profiles. 
The disk integration algorithm approximates 
the star as a 500$\times$1000 grid. The spectrum was computed at steps 
of 0.01\AA. We should note that our 
adaptation of Kurucz model atmospheres with normal {\bf abundances 
does not} include the effect of enhanced backwarming through the atmosphere
due to the overabundances of the Fe group ions. 
Fortunately, non-LTE effects appear to be small for most UV metallic
spectra of B stars with $T_{\rm eff}$ $<$ 25,000\,K (e.g., Smith, Sterken,
\& Fullerton 2005). 

  Our goal in this analysis will be to assess the effects of an 
absorbing magnetospheric disk on UV lines in the star's spectrum as the disk
moves across and off the star's surface. To simulate the signatures of this 
structure on the composite spectrum, we used the radiative transfer
program {\it CIRCUS} (Hubeny \& Heap 1996, Hubeny \& Lanz 1996) to compute
strengths of absorption or emission components of lines in a circumstellar
medium from input quantities such as disk temperature, volume
density, column density, areal coverage factor (the fraction of the star
covered by the disk at maximum occultation phase), and microturbulent 
velocity. Although {\it CIRCUS} can accommodate as many as three separate 
circumstellar ``clouds" with as many independent parameter sets, in this
paper we assume the disk to be homogeneous. 
% (The sole exception to this is for the computation of line-blended UV 
% resonance lines, for which we used a large turbulent velocity.)
In its solution of the radiative transfer, {\it CIRCUS} allows the user to 
compute the line flux with the line source function set either to zero
or the Planck function at the local temperature. This feature can be 
used to compute the disk spectrum in full non-LTE (scattering approximation) 
or LTE modes, respectively. {\bf We used the scattering approximation for our
analysis of UV resonance lines.}

  Other key input parameters include the disk volume density 
and turbulent velocity.  The volume density enters
only through the ionization of the species, which is 
mainly determined by the temperature. Our input density 10$^{11}$ 
cm$^{-3}$ was the order of magnitude determined for hot Bp star disks 
by SG01. This parameter was used only to set the ionization equilibria for 
the elements responsible for the resonance lines. However, we note that 
these equilibria are mainly sensitive to temperature.
Except for the absorptions of the Si\,IV, C\,IV, and N\,V
resonance lines, the overwhelming majority of the UV metal lines in 
our computations are optically thin. Since the spectral signatures then 
depend on the volume of absorbers, a degeneracy arises between two 
geometrical factors of the disk: its area and column length along the 
observer's line of sight. For convenience we have performed our 
spectral analyses, excluding the occultation models discussed in 
$\S$\ref{modl}, by assuming that the coverage factor of the disk
is unity when it passes in front of the star. Our analysis, given our
choice of $i$ and $\beta$, will validate this assumption.

  The final ingredient of our models is the treatment of line of 
sight velocities. 
Because the magnetosphere co-rotates with the star's surface, 
no differential velocities exist along the line of sight to the star
during occultation (transit). SG01 found
that turbulent velocities in the range of 0\,--\,20 km\,s$^{-1}$ resulted 
in similar absorptions of ensembles of iron lines. SG01 also found
a broadening of the Al\,III resonance lines of at least 25  km\,s$^{-1}$
at occultation phases $\phi$ = 0.0 and 0.5.  
Based principally on this result, for the analysis of the disk lines, we have 
adopted an arbitrary value of 20 km\,s$^{-1}$, or about twice the local 
sound speed. 

%FIGURE 1 old
   \begin{figure}[t]
   \centering
   \includegraphics[width=6.5cm,angle=90]{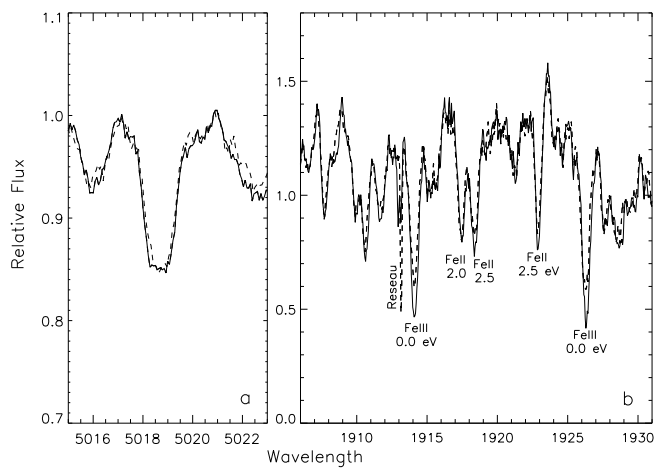}
\caption{{\it Panel a:} A comparison of two observations 
of the Fe\,II $\lambda$5018 line region obtained 
at during nonoccultation $\phi$ = 0.90 (2001 December 02; dashed line) and at 
the occultation phase, $\phi$ = 0.02 (2000 February 12; solid line). {\it Panel
b:} A comparison of high-resolution {\it IUE} (unsmoothed) spectra centered 
on a wavelength region that contains a mixture of strong Fe\,II  and Fe\,III 
lines. {\bf The fluxes in {\it a} and {\it b} are relative, or are 
given in c.g.s.  units multiplied by 10$^{10}$, respectively.   }
The dashed line spectra represent the photospheric spectrum while 
the solid line depicts the spectrum at occultation phases. 
}
         \label{fe2fe3}
   \end{figure}

\section{Basic picture}

\subsection{The Goldilocks absorption effect on iron lines} 
\label{goldy}

  In addition to the modulation of the resonance lines, one of the 
phenomena that drew our attention to 36\,Lyn was the contrasting 
behavior of its iron lines in the optical and UV spectral regions.
The two panels of Figure\,\ref{fe2fe3} show the different behaviors 
of strong optical Fe\,II and UV Fe\,III lines during occultation and
nonoccultation phases. In Fig.\,\ref{fe2fe3}a the line
core of the Fe\,II $\lambda$5018 at the occultation phase $\phi$ = 0.97 
(solid line)
is no deeper than during the nonoccultation phase, $\phi$ = 0.82. It is
possible that weak bumps occur on the latter profile. 
These are likely to be signatures of an inhomogenous surface distribution
of Fe on the star's surface, as found in Paper\,1.
In contrast to the optical lines, Fig.\,\ref{fe2fe3}b 
shows that several low-excitation Fe\,II and Fe\,III 
lines in the $\lambda$1900 region deepen significantly during occultation 
phase. This is again consistent with the strengthenings of low-excitation 
lines during disk occultation of the hot stars studied by SG01. The
strongly contrasting different behaviors of the optical and UV lines
(the former responds to surface inhomogeneities and the latter to 
circumstellar conditions) may seem surprising at first. However, they can 
be understood after computing the opacity spectra for physical conditions 
we will find in the disk. Taking $T_{disk}$ = 7,500\,K and  microturbulence
$\xi$ = 20 km\,s$^{-1},$ we find that the Fe\,II $\lambda$4923 and 
$\lambda$5018 lines in this disk medium have opacities of about 
4$\times$10$^{-23}$ cm$^{-2}$ atom$^{-1}.$ By comparison, the opacities 
of the UV iron lines depicted in Fig.\,\ref{fe2fe3}b are typically 50$\times$ 
higher than this value. These values fall 
in a region of parameter space for which the disk absorptions of 
optical lines are practically invisible. In contrast, the optical 
depths of the UV lines are near unity, and thus they respond
rather sensitively to changes in column length. Because we are in a 
window for which the opacities are just right, we 
call this a ``Goldilocks" absorption effect. Thus, for Bp stars with higher 
effective temperatures, we might expect various surface 
and disk conditions to cause competing effects on the optical and UV 
metallic line profiles. Abundance patches along the magnetic equator
will cause high-excitation lines (those usually visible at optical
wavelengths) to strengthen as these regions cross the observer's meridian, 
whereas low-excitation lines would strengthen at disk occultation
phases. The result could be confusing for a panchromatic abundance 
analysis if the presence of a disk were not taken into account.
For {\bf stars cooler} than 36\,Lyn winds will be very weak, 
and we expect surface inhomogeneities due to diffusion to dominate all 
the lines.

\section{Spectral Analysis of Line Absorption}

Our analysis of the disk properties begins by examining the
ionization of iron atoms in the disk, first by comparing specific strong
lines in the spectrum and then by assessing their group effects in 
differential spectrophotometry.

\subsection{Disk diagnostics from lines of iron and low-ionization atoms}
\label{fedisk}

\subsubsection{Analysis of individual lines}
\label{analind}

  The ionization equilibirium of iron shifts quickly from Fe$^{1+}$ to 
Fe$^{2+}$ in the temperature range 7,000\,K\,--\,8,000\,K. One can 
exploit this sensitivity 
by comparing the increases of Fe\,II and Fe\,III lines 
during occultation. A sample of these lines is shown 
in Fig\,\ref{fe2fe3}b. We have computed {\it CIRCUS} models to simulate 
these strengthenings in detail 
in order to determine both a mean temperature and column density of the 
disk. By matching the {\it ratio} of Fe\,II to Fe\,III strengthenings during
occultation phase, we found that the temperature is tightly constrained 
to values of 7,500$\pm{500}$\,K. Our internal errors in this determination 
have been doubled in order to take into account any potential effects 
of non-LTE on iron line formation. The same models, but now fitting to the 
{\it mean} increase in iron-line strengthenings, led to a less secure 
estimate of the column density, 
3$_{-1.5}$$\hspace*{-.25in}^{+3}$$\hspace*{.1in}\times$10$^{20}$\,cm$^{-2}$. 
We estimate
the error on this parameter to be a factor of three, based on uncertainties 
in disk coverage factor, the assumed high Fe abundance of the disk, 
microturbulence, and our oversimplification of the homogeneity of the disk. 
For low stages of ionization, the ionization potentials of aluminum are similar 
to iron. This fact encouraged us also also to model the variations of
other resonance lines of this element, such as Al\,II $\lambda$1670, and 
Al\,III $\lambda$1855, $\lambda$1863. As before, we exploited the rapid 
ionization shift from Al$^{1+}$ to Al$^{2+}$ in the temperature range
7,000\,--\,8000\,K. Again, we found a best fit with 7,500$\pm{500}$\,K.
The fits to the mean strengthenings of the line during occultation phases
led to a column density of 1.5$\times10^{20}$ cm$^{-2}$. We estimate a
factor of two {\bf for the} error in {\bf the latter value.} The column 
density determination from aluminum lines is well within the {\bf error range}
found from individual Fe\,II and Fe\,III lines.

   We have similarly attempted to constrain temperatures from silicon 
lines arising from neighboring ionization stages, but this has proven
difficult. In the case of silicon, only two unblended Si\,III lines are
strong enough to be visible. In addition, the mean strength of several
apparent ``Si\,III" lines is overestimated because of 
close blends with iron lines. These tend to drive the apparent abundance
of Si$^{2+}$ well above optically-derived values. The absence of variation 
in the forbidden Si\,III resonance line at $\lambda$1892 (which would 
otherwise be useful for analysis of extended low-density media)
can be used only to rule out a temperature above 9,000\,K and/or a column 
density above 3$\times10^{21}$ cm$^{-2}$. 
The Si\,III resonance line at $\lambda$1265 has a broad saturated core. 
At occultation phase $\phi$ $\approx$ 0.0, the core strengthens and broadens. 
We can duplicate this strengthening only for $\xi$ $>$ 20 km\,s$^{-1}$ in 
our models. This is true to even a greater extent for the C\,II 
$\lambda\lambda$1335-6 doublet, which, for a good fit of the line core and 
strength variations, requires large $\xi$ values of 50 km\,s$^{-1}$.

\subsubsection{Analysis of the ``Iron Curtain" lines}
% \label{fecurtn}

\begin{figure}[t]
 \centering
   \includegraphics[width=6.5cm,angle=90]{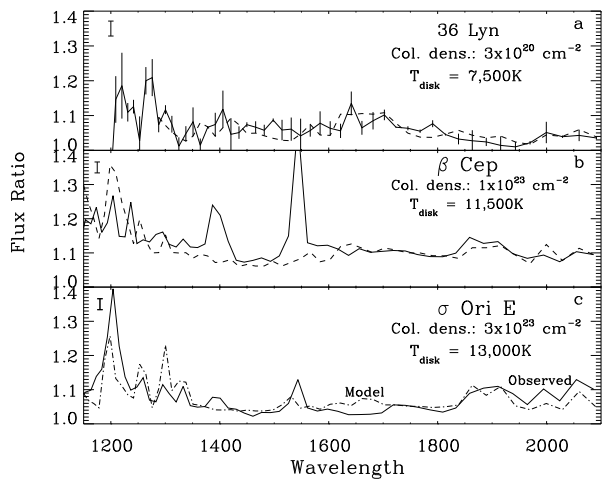}
\caption{Observed (solid line) and computed-plus-1.0 (dashed line) ratios of  
disk Fe-absorption spectra for 36\,Lyn, $\beta$\,Cep, and $\sigma$\,Ori\,E.  
Spectra for 36\,Lyn are the same as those used for Fig.\,\ref{fe2fe3}.  
The models indicated assume a full disk occultation of the star and a
value $\xi$ = 20 km\,s$^{-1}.$ The temperature used in this fit was
taken from the fittings discussed in $\S$\,\ref{analind}.}  
\label{fecurt}
\end{figure}

Following the methodology of SG01, we used {\it CIRCUS} and {\it SYNSPEC} 
to fit the strengthenings of the absorption lines during disk occultation.  
This analysis was performed by first summing all fluxes within each 
echelle order of a {\it IUE/SWP} spectrum and then computing the mean 
absorption in this order for all spectra observed at similar phases. 
The division of occulted by nonocculted phase {\bf spectra}, minus one, gives 
the fractional absorption in an UV spectrophotometric representation of th
observations. This divided low-resolution spectrum can be compared directly 
with the computed photospheric spectrum divided by the computed 
spectrum of the star absorbed by the model disk.
For the disk spectrum of 36\,Lyn, the overwhelming majority of lines are 
optically thin. This fact introduces a degeneracy between the disk
areal coverage (effectively, the disk ``height") and the column density.
Since the UV opacity is almost entirely due to Fe-group 
elements, a second fitting degeneracy occurs between column density 
and iron abundance. We also note
that the value $\xi$ = 20  km\,s$^{-1}$ produces a 
wavelength-averaged absorption that is 35\% higher than the mean absorption 
produced by $\xi$ = 10 km\,s$^{-1}$. These uncertainties enter
heavily into the error budget for the column density given above.
Using these parameters, a disk temperature of 7,500\,K, and an assumed
photospheric value of $T_{\rm eff}$ = 13,000\,K, we modeled the UV 
spectrophotometric data. The comparison of this {\it CIRCUS} 
model with the observations is given in Fig.\,\ref{fecurt}a. 
In Figs.\,\ref{fecurt}b and \ref{fecurt}c we exhibit the comparisons 
for $\beta$\,Cep and $\sigma$\,Ori\,E, which are taken from SG01. 

To give the reader an idea of strengths of ``typical" lines in the UV 
spectrum, we note that for 7,500\,K the median cross section for lines 
with wavelengths
above $\lambda$1300 is 20 times stronger than the underlying hydrogenic
opacity (2$\times10^{-23}$ cm$^{-2}$ per atom). Only 2.5\% of lines in 
the wavelength range $\lambda\lambda$1300--2100 are optically thick 
for our reference column density of 3$\times10^{20}$ cm$^{-2}$. 
Because of the degeneracy between coverage and column density
one can match the mean levels of absorption by increasing the disk 
column density by factors of 6 and 30 while decreasing its coverage 
area by factors of 10 and 100, respectively.

%FIGURE 2 old

  Bjorkman, Bjorkman, \& Wood (2000) and Smith (2001) have pointed 
out that for spectra of circumstellar disks around B stars, aggregates of 
strong lines in selected wavelength intervals provide temperature
diagnostics. The character of these aggregates shifts through changes in 
ionization. For example, a well-known ``Fe\,III bump" appears in the 
disk temperature range 10,000\,--\,16,000\,K. 
This bump is visible in the spectrophotometry of the hot Bp stars analyzed 
by SG01, but it does not appear in the 36\,Lyn data. However, 
Fig.\,\ref{fecurt}a {\bf reveals 
a just detectable bump \bf in} the range $\lambda\lambda$1550-1750.
Inspection of our synthesis results shows that this feature is composed
almost entirely of optically thin Fe\,II lines, consistent with our
assumptions. Putting all these facts together, and again using 
a coverage factor of 100\%, an iron abundance of +1.04 
dex, $T_{disk}$= 7,500\,K, and $\xi$ = 20 km\,s$^{-1}$, our {\bf fitting to
the} mean absorptions in Fig.\,\ref{fecurt}a leads to a column density
of 3$\times10^{20}$ cm$^{-2}$. We also notice that the column density
is at least a factor of 100 lower than those found in disks of the hot Bp 
stars $\sigma$\,Ori\,E and $\beta$\,Cep (SG01). According to a preliminary
analysis, the column density for the disk of another B8p star, HD\,21699, 
is probably even lower, suggesting that the disk column tends to increase
monotonically with effective temperature among magnetic Bp stars. Yet,
there is no trend in disk parameters with the stars' field strengths.
 
 The agreement of the disk temperatures inferred from individual Fe
and Al lines on one hand and the spectrophotometric curtain absorptions on
the other {\bf hand} is important in justifying our geometric results below
that the inner region of the disk is largely evacuated. Recent 
radiative equilibrium models of circumstellar disks of Be stars
assume a slow outward flow of matter such that the density decreases 
outward with radius. These models predict a local {\bf temperature} 
minimum in this disk component. For 36\,Lyn this temperature should be
$\sim$6,000\,K. However, our models indicate a strengthening of many
Fe\,II lines that is not observed. This fact argues against the presence 
of a substantial contribution from a low temperature component to the disk.

\subsection{Ultraviolet resonance lines}

\subsubsection{Basic Characteristics of the resonance doublets} 
\label{uvres}
% Delta EWs of CIV doublet:  0.550, 0.440 AA
% Delta EWs of SIIV doublet:  0.173, 0.155 AA
% Delta EWs of NxIV doublet:  0.065, 0.040 AA

\begin{figure}[t]
   \centering
   \includegraphics[width=6.5cm,angle=90]{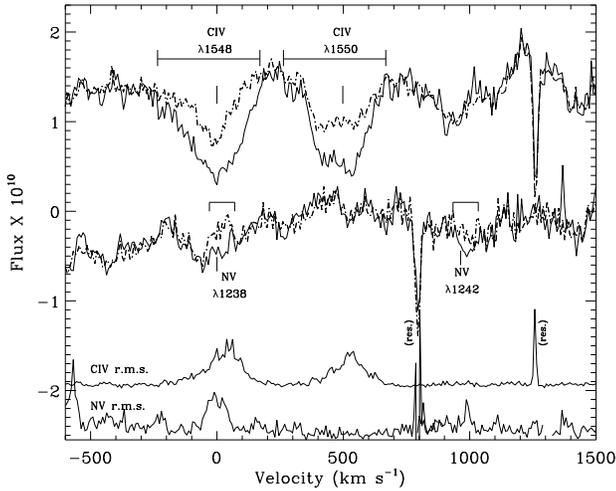}
\caption{Maximum (solid line) and minimum (dot dashed line) absorption spectra for 
the C\,IV and N\,V lines in 36\,Lyn (upper plots) and their renormalized
{\bf difference} r.m.s. spectra {\bf of all spectra}
(lower plots). The spectra represented are those in Figs.\,1b and 2c. The
velocity reference is zero for the rest frames of the $\lambda$1548 and 
$\lambda$1238 components of the C\,IV and N\,V doublets, respectively. 
Instrumental ``reseaux" are indicated.} 
\label{c4rms}
\end{figure}

  To determine the characteristics of the C\,IV, Si\,IV, and N\,V spectra,
we first plotted these lines' equivalent widths with respect to
rotation phase. This was accomplished by using the phase zero point and 
period given in Table\,1. We used these to form {\bf two average spectra: one
taken near mid-occultation phase and the other near the quadratures when
the disk is seen face-on.}
The two resulting spectra were taken from
{\it IUE} observations SWP\,27038, 27052, 27070, 27091, 
31996, 32014-5, 32947, 32971-2 and SWP\,32024, 32057, 32990, respectively. 
Figure\,\ref{c4rms} compares these spectra for the C\,IV and N\,IV doublets
The bottom of this figure gives a sense of the statistical variations 
through the r.m.s. spectra for these lines from the available data.
Detailed modeling of the resonance lines in  both the photospheric 
(actually ``disk face-on") and disk-absorbed spectra
with {\it SYNSPEC} shows that the C\,IV and Si\,IV lines suffer blends.

%%FIGURE 3 old

% FIGURE 4 old
   \begin{figure}[h]
   \centering
   \includegraphics[width=6.0cm,angle=90]{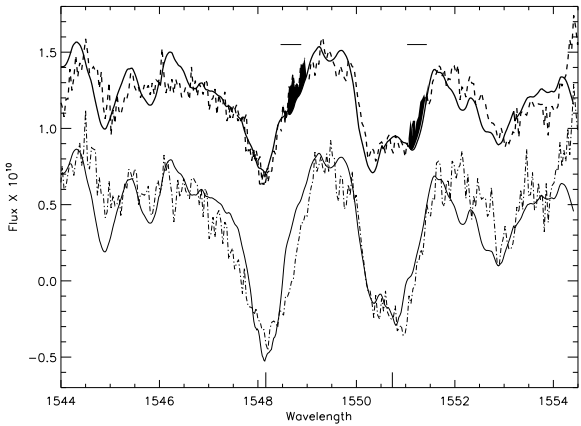}
\caption{The observed maximum (dot-dashed line) and minimum (thick dashed 
line) absorption spectra for the C\,IV lines in 36\,Lyn. Corresponding spectra 
computed for the photosphere and during disk occultation (solid lines)
are shown. The maximum and 
disk-absorbed spectra are shifted -0.7 units for clarity.
The dark shading, interpreted as emission, is the difference between the
minimum absorption and photospheric profiles.}
\label{c4mxmn}
\end{figure}

   Figure\,\ref{c4mxmn} exhibits examples of the computed and observed 
template spectra of the C\,IV doublet at quadrature and mid-occultation 
phases. We have offset the disk-absorbed spectra downward for clarity. 
As this figure shows, the photospheric spectra can be modeled well using
the solar C and the Fe-rich abundances specified in Paper 1.  The 
disk-absorbed spectra of C\,IV can also be well modeled, with one
stipulation, using a disk temperature $T_{disk}$ = 25,000\,K and the column 
density given in $\S$\ref{analind}. {\bf We stipulate that our residual} 
mismatches probably result from our use of a disk microturbulence, 
20 km\,s$^{-1}$, that is probably much less than the turbulence in at
least some regions of the disk. Subsequent trials showed that values of 
at least 50 km\,s$^{-1}$ are required. High turbulence values (if it 
is a true microturbulence) were also noted by SG01 and were also reported
for the our modeling of 36 Lyn's C\,II doublet profiles in $\S$\ref{analind}.
The mismatches can be seen in the broad core of the $\lambda$1548 line in 
Fig.\,\ref{c4mxmn} (lower solid line). The mismatch is not so noticeable in 
the $\lambda$1550 feature because it is a semi-resolved blend of the C\,IV
$\lambda$1550 component and Fe\,II $\lambda$1550.3; the added turbulence
is not as large as the 0.5\AA\ separation between the two $\lambda$1550
lines. Another interesting mismatch is the series of low-excitation Fe\,II
lines in the regions $\lambda\lambda$1544--6 and $\lambda\lambda$1552--3
for the disk-absorbed spectrum. Our simulations suggest that a higher 
microturbulence would ameliorate these mismatches. However, these
high turbulences, for which we have so far no physical explanation,
are not otherwise important to this analysis. 

  The measured equivalent width differences of the C\,IV $\lambda$1548
and $\lambda$1550 components are 550\,m\AA\ and 440\,m\AA,~ respectively.
The corresponding maximum and minimum equivalent widths of the Si\,IV 
$\lambda$1394 and $\lambda$1403 lines are 173\,m\AA\ and 155\,m\AA,~
and for the N\,V doublet 65\,m\AA\ and 40\,m\AA,~ respectively.
The C\,IV and Si\,IV doublet ratios are close to unity. 
{\bf Therefore, we} conclude that these lines are optically thick.

  The small shaded regions in Fig.\,\ref{c4mxmn} are possible
emissions due to the disk in the plane of the sky. These
can be expected to contribute line-scattered flux in the
line of sight to observer at this phase. The emissions measured from
the face-on (observed) minus computed (photosphere only) spectrum
are -31\,m\AA\ and -15\,m\AA,~ which corresponds to 5\% (${\pm 5}$\%),
of the total difference. Given the errors on these quantities, they
should not be overinterpreted. However, we may safely say that any
emission contributed by the disk in these lines is small or negligible.

% Table 2
\begin{table*}[ht!]
\begin{center}
\begin{tabular}{cccc|ccc}  \hline\hline
N\,V ($\lambda$1238)~~~~~ &$T_{disk}$ & Col. Dens. & $\tau_{line}$& $T_{disk}$ & Col Dens.&$\tau_{line}$\\ 

                 &  &  (Abs.)   &              &  &  (Emis.)   &      \\
$T_{max}$ & 53,000~ & 1.5$\times10^{19}$& .001 & 69,000~  & 6$\times10^{17}$&.001\\
$T_{mid}$ & 43,000~ & 1$\times10^{18}$ & .05 & 50,000~ &5$\times10^{13}$&.005 \\
$T_{min}$ & {\it 33,000$^*$} & {\it 1$\times10^{10}$} & {\it .03} & {\it 30,000$^*$}& {\it 6$\times10^{17}$} & {\it .02} \\
\hline
~C\,IV ($\lambda$1548)~~~~~ &$T_{disk}$ & Col. Dens. & $\tau_{line}$& $T_{disk}$ & Col Dens.& $\tau_{line}$ \\

                 &  &  (Abs.)        &              & &  (Emis.) &     \\
$T_{max}$ & 30,000~  & 5$\times10^{19}$& 0.3 & 50,000~  & 2$\times10^{17}$&0.03\\
$T_{mid}$ & 28,000~ & 1$\times10^{18}$ & 3 & 30,000~ &1$\times10^{14}$& 0.003 \\
$T_{mid}$ & &  &  & 25,000~ &  2$\times10^{16}$ & 0.03 \\
{\it $T_{min}$} & {\it 25,000$^*$} & {\it 4$\times10^{19}$} & {\it 12} & {\it 20,000$^*$}  & {\it 6$\times10^{17}$} & {\it 3} \\
\hline
\end{tabular}
\begin{tabular}{cccc|ccc}  
Si\,IV ($\lambda$1394)~~~~~~ & $T_{disk}$ & Col. Dens. & $\tau_{line}$ & 
~~~~~~~& (No emission)   & ~~~~~~  \\
            &    &  (Abs.)   &   & & &                 \\
$T_{max}$ & 30,000~  & 7$\times10^{18}$ & .004 & & &  \\
$T_{mid}$ & 20,000~  & 1$\times10^{14}$ & .01  & & & \\
{\it $T_{min}$} & {\it 15,000$^*$} & {\it 4$\times10^{17}$} & {\it 4} & & & \\
\hline
Si\,III ($\lambda$1206)~~~~~ &$T_{disk}$ & Col. Dens. & $\tau_{line}$  & 
 &   (No emission)  & \\
            &    &  (Abs.)   &    &     &   &               \\
$T_{max}$ & 20,000~  & 1$\times10^{18}$ & .03 \\
$T_{mid}$ & 15,000~  & 1$\times10^{18}$ & 1  \\
{\it $T_{min}$} &  {\it 10,000$^*$} & {\it 1$\times10^{19}$} & 22    \\ 
 \hline
\end{tabular}
\caption{\label{tab2}{Derived disk temperatures (K), column densities (cm$^{-2}$), and $\tau_{line}$ for model fits for UV resonance lines.}} 
\end{center}
\end{table*}

\subsubsection{Temperature and column density results for the resonance lines}
\label{cdtemp}

  Our modeling of the equivalent widths of the resonance lines is 
most sensitive to the estimated temperature and column density. Therefore, 
other arguments must be used to constrain these values. For a sample derived 
from absorptions of just a few lines, a temperature solution is generally 
poorly constrained, and so the argument is frequently made that the most 
likely temperature is that which results in maximum absorption (or emission), 
or equivalently, minimum column density. Moreover, the optical thickness 
requirement on the C\,IV and Si\,IV absorptions can be used to constrain 
the region of temperature and density space. 
In Table\,2 we present disk temperatures and column densities required to 
fit the absorption equivalent widths measured for the stronger component of
the N\,V, C\,IV, Si\,IV and Si\,III resonance doublets. 
The column densities displayed for the emission cases in the table are
somewhat arbitrary and are intended to show $\tau$ scales for the lines. 
In these computations we have utilized photospheric abundances and coverage 
factors of unity. Because we do not have foreknowledge of the line 
excitation temperatures in the disk, we computed the strengths 
both in emission and absorption for a range of temperatures and evaluated
them to determine temperature limits. We also 
searched for ``optimal " values of the emission/absorption temperatures. 
Optimal temperatures are
those for which lines are formed most efficiently and thus require 
minimum column densities per unit $\tau$. Lower and upper temperature 
limits were likewise evaluated by assuming that the gas responsible for 
absorptions might have column densities as large as $\sim$10$^{19}$ 
cm$^{-2}$. This is equivalent to a few percent of the column lengths 
derived in $\S$\ref{fedisk}. Even if this assumed column value were too low, 
the ionization fractions at these limits drop off quickly with temperature. 
Thus, a smaller fraction would produce little change in our temperature 
ranges. The values given in the ``emission" columns of Table\,2 were 
obtained by assuming that, as for C\,IV in the previous section, 
they are equal to $\approx$5\% of the range of the equivalent width 
variation.  From the computed line cross sections determined in 
these calculations, we find that the {\it emission} column densities 
are low enough for the lines to be optically thin, contrary to their
optical thickness when observed at occultation.

  One may appeal to the formation process of the
``super ions" to narrow the temperature ranges still further from those 
given in Table\,2 because the resonance lines are produced by an 
istropic process, namely scattering. Then because the line photon
addition or removal is caused by the same disk ions (seen from
different viewing angles at these phases), one expects the column 
density determined for the two cases to be roughly the same. 
This expectation is not met in the C\,IV and N\,V column densities 
{\bf given} in Table\,2. In fact, the best absorption column values
are at least an order of magnitude higher than for the emissions.
This fact suggests that one of the two temperature limits, 
and not the optimal one, offers the best solution.
According to the higher optical depths in the C\,IV lines indicated by the
measured doublet ratios, this limit is likely to be the lower one. 
We suspect it is the lower of the two limits for the N\,V lines too.
This reasoning suggests that the true temperature range derived for
these lines is smaller than the values tabulated. Based on
these considerations, we represent in Table\,2 our estimates of
most realistic temperatures by asterisks and italics. 
In cases where a temperature limit seems to 
be indicated, we have chosen the lower one.

\subsubsection{Variation of the resonance lines with phase}

  To quantify the behavior of the resonance lines with phase, 
we have computed line strength indices of 36\,Lyn by 
computing the ratio of fluxes within narrow velocity limits around the 
line center(s) to the residual fluxes in the net (unrectified) spectra 
contained in the parent echelle order. This technique is not subject to 
errors of continuum rectification and placement. The measurement limits
were set to $\pm{65}$ km\,s$^{-1}$ from line center, or just
beyond the rotational limits on the profile. A comparison of indices extracted 
from both these broader and narrower limits for C\,IV showed that they 
followed the same form, but the broader limits decreased the apparent errors.

\begin{figure}[t]
   \centering
   \includegraphics[width=6.5cm,angle=90]{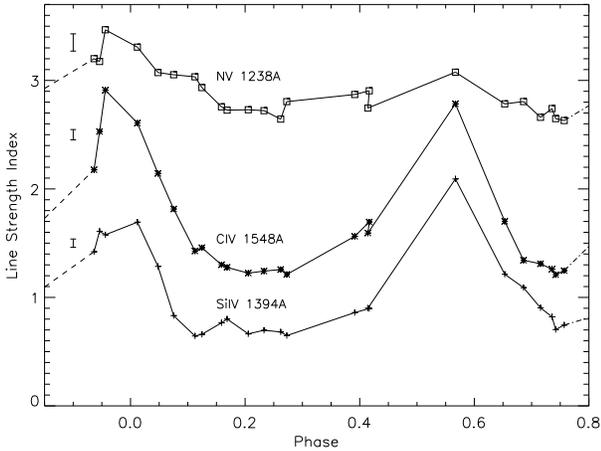}
\caption{
{\bf Line strength indices} of the N\,V $\lambda$1238,
%(squares),  
C\,IV $\lambda$1548, 
%(asterisks), 
and Si\,IV $\lambda$1394 
%(pluses) 
lines for 36\,Lyn with magnetic/rotation phase. The curves have been 
offset for clarity.}
\label{ewphs}
\end{figure}

Figure\,\ref{ewphs} depicts the line strength indices with phase 
for C\,IV, {\bf Si\,IV,} and N\,V. 
The curves exhibit a narrow primary peak centered 
near $\phi$ = 0.00, which for 36\,Lyn corresponds to the rotation
of the South magnetic pole onto the receding limb of the star 
and an underresolved ``secondary" peak occurring near 0.55 cycles. 

%FIGURE 5 old

 Note in Fig.\,\ref{ewphs} that the relative C\,IV absorption peaks have 
nearly the same heights. However, for Si\,IV the second peak is stronger,
while it is much weaker for N\,V,  as indeed we find it is also for 
H$\alpha$ ($\S$\ref{alffit}) and the less excited ions of
Si\,III, Al\,III, and Fe\,III (not shown). 
The differing ratios suggest that the range of ionization temperatures 
of gas within the sector causing the first occultation is narrower than
for the secondary occultation sector.  {\bf MacFarlane, 
Cohen, \& Wang (1994) have pointed out that models of photoexcitation
of ions by soft X-rays predict a broad distribution of excited ions. 
This does not seem to occur in the primary occultation segment, suggesting
that this segment (and probably the secondary one as well) is excited by
collisions and not photoionizations.}

\section{The line strength-phase curves of the resonance and H$\alpha$ lines}
\label{modl}

\subsection{The models}

   The morphologies of the {\bf line strength}-phase curves, such as those shown
in Fig.\,\ref{ewphs} and the H$\alpha$ curve shown in Fig.\,4 of Paper\,1,
offer the possibility of determining geometrical properties of the
disks of Bp stars. Although the disk column density 
is relatively low, the ratio of the absorptions of the resonance doublet
members indicates that these features are optically thick. This means that 
the absorption process can be recast by merely computing the occulted area
of the star. We wrote a program to compute this absorption by considering 
a cylindrical opaque ring or disk with origin coinciding with the star's
center. Its tilt with respect to the rotational {\bf equator} is fixed by the 
magnetic obliquity $\beta.$ In our 
formulation all light from the star is occulted by this structure. 
Physically this means that the source function relative
to the background stellar photospheric emissivity is negligible.
This could mean, for example, that the radiative transfer in the
lines is dominated by resonance scattering. However, it is also 
possible that a similar absorption profile can result if the line is 
formed in LTE. The key requirement in either case is that the emergent
flux from the slab with changing viewing angle remains constant during
the occultation event. 

In our construct the magnetic and rotational obliquities may be set to 
arbitrary values. We adopted the values $i$ = $\beta$ = 90$^o$, 
which are consistent with the limits found in Paper 1.  Thus, our
basic model of the ``disk" reduces to an opaque cylindrical ring oriented
perpendicular to the star's rotational axis. As the star rotates, one or
the other of the disk segments rotates in front of the star, except at 
quadrature phases 0.25 and 0.75 when the magnetic poles cross the center
of the disk.  The free parameters of our opaque cylinderical model consist 
of inner and outer disk radii, given by $r_i$ and $r_o$, and a disk semiheight 
$h,$ out of the equatorial plane, all of which are expressed in units of 
R$_{*}$.  Because the circumstellar 
disk is assumed to be axisymmetric and centered at the center of the star,
one need only integrate over half the star's area. We used a star disk 
grid of 2000$\times$1000 pixels and integrated the data over 1$^o$ 
phase intervals.  Because of the appearance of an ingress feature discussed
below, we eventually modified
our program to approximate the effects of an optically thin extension 
out of the magnetic and disk plane. In this second class of models, the 
extension was approximated by a translucent absorption that falls off
from an initial value of 1/$e$ at the edge of the opaque component. 
This absorption is parameterized by a second scale height $h_{S}.$ This
component is layered over/under the opaque components, making it extend
further out from the plane.

%FIGURE 6 old
   \begin{figure*}[t]
   \centering
   \includegraphics[width=11.5cm,angle=90]{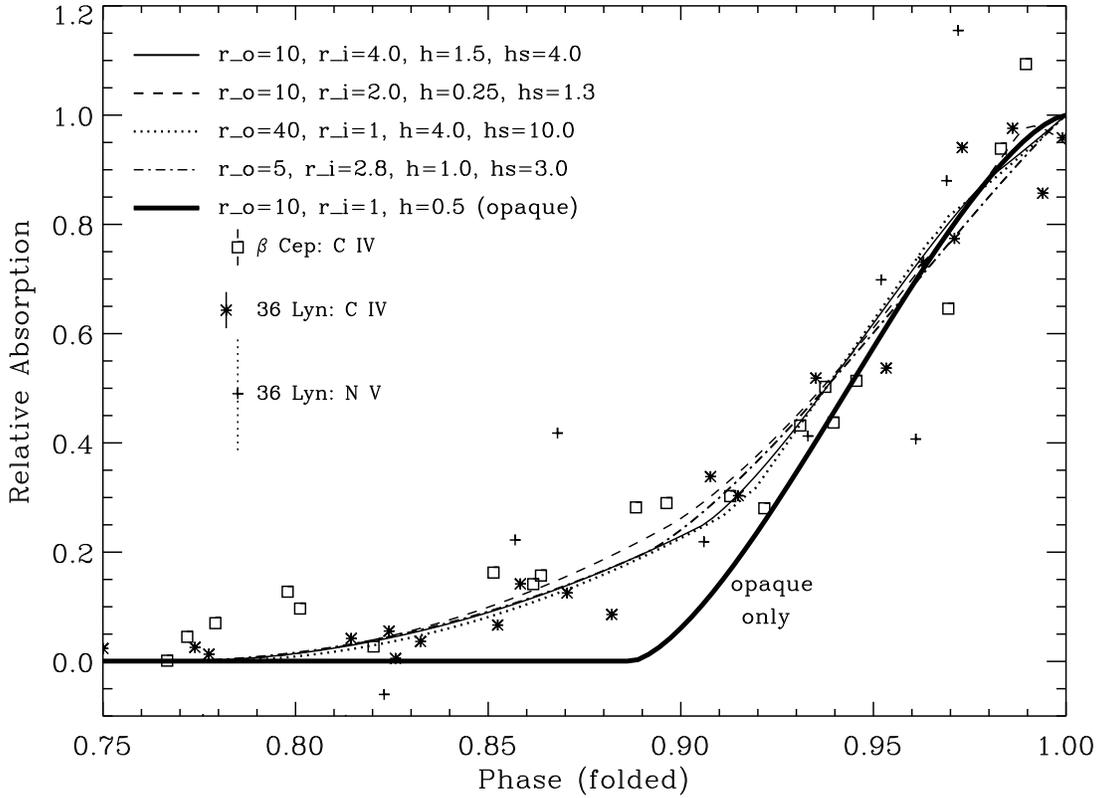}
\caption{Fits to excess absorption with phase of the C\,IV and N\,IV 
resonance lines of 36\,Lyn and C\,IV lines of $\beta$\,Cep at primary 
occultation. The data are taken 
from Fig.\,\ref{ewphs} and SG01, respectively, and folded around $\phi$ = 1.0
(disk viewed edge-on). Error bars are indicated on the data symbol annotations.
The various coded lines are fits from the 
two-component model and, where indicated, single opaque component model 
% for the disk absorption ($r_o$ = 20R$_*$, $r_i$=3R$_*$, $h$=1R$_*$). 
for the disk absorption ($r_o$ = 10R$_*$, $r_i$=1R$_*$, $h$=0.5R$_*$). 
For convenience all fits indicated
were computed for the simple geometrical case of the star viewed 
equator-on and with a magnetic inclination of 90$^o$. 
}
\label{c4phs}
\end{figure*}

  A weakness of our formulation is interpretational: one does not
know where along the line of sight the absorption occurs at a given viewing 
angle. Despite this limitation, our approach offers the advantage of not
tying the results artificially to a completely opaque disk geometry. A
secondary, minor weakness in our picture is that the H$\alpha$ opacity appears 
not to be infinite ($\tau$ $\sim$ 5 in our CIRCUS models). 
 
  Before we start the geometrical analysis, we ask what 
limits apply to the extent of the radial disk. We can estimate
this {\bf extent} by computing 
the Alfv\'en radius R$_{A}$, which is the radius at which the magnetic 
and wind energy densities are equal. This is given by the simple 
relation:

\begin{equation}
 ~~~~~~~~~~~~~~~~~~~~~~~~~~~~~~ R_{A} = \eta^{\frac 14},
\end{equation}

\noindent where  $\eta$, or the ratio of magnetic to 
wind energy densities (uO02), is given by:

\begin{equation}
~~~~~~~~~~~~~~~~~~~~~~ \eta_{*} = B_*^2 R_*^{2}/(\dot{M} v_\infty)~.
\end{equation}

\noindent Here $B$ is the mean surface field, $R_{*}$ the stellar radius, 
$\dot{M}$ the mass loss rate, and $v_{\infty}$ is the terminal wind velocity.
From Paper\,1, we take $B$ = 3\,kG and R$_{*}$ = 3.4R$_{\sun}$. BM97
considered a range of mass loss values 
10$^{-10}$\,--\,10$^{-11}$ $M_{\sun}$\,yr$^{-1}$ for the active X-ray
star IQ\,Aur ($T_{eff}$=13,000--16,000\,K). 
36\,Lyn is undetected in the X-ray regime, so 
we will assume $\dot{M}$ = 10$^{-11}$ $M_{\sun}$\,yr$^{-1}$. Effects 
of the wind are not detectable on the resonance lines, so
we will assume an arbitrary value v$_{\infty}$
= 600 km\,s$^{-1}$ (BM97 assumed 800 km\,s$^{-1}$).
These values lead to a value $\eta$ $\approx$ 10$^{7}$. Then, an application
of equation\,(1) yields R$_{A}$ $\approx$ 50 R$_{*}$. Although
these simple relations may or may not scale precisely in a high
$\eta$ regime, we can still be assured that the Alfv\'en radius for 36\,Lyn
is very large. We will be at liberty to consider disk radii of a few
tens of stellar radii.

\subsection{Fitting of UV resonance line variations with phase}
 
  The line strength curves of the resonance lines of 36\,Lyn shown in
Fig.\,{\bf \ref{ewphs}} have three general features: (1) a peak with central 
maximum near $\phi$ = 0.0, (2) a quick drop-off to a narrow core (a 
half-width half-maximum of 0.06 cycles), and 
(3) a tapered ingress/egress ``tail" extending from phase {\bf 1.00} to 
$\approx$ 0.82 and 0.18 ({\bf 1.18}).  To study these curves further, we folded 
the C\,IV and N\,V equivalent widths measured around the occultation 
midpoint phases. 
For reference, we did the same for the C\,IV data of $\beta$\,Cep. 
For both stars the estimated phase of maximum absorption is not exactly 
at 0.00. In the interest of coaligning the peaks of the two stars and 
{\bf comparing the morphologies,} we shifted the phases slightly to
remove this difference. The Fig.\,\ref{ewphs} data manipulated in this 
fashion are shown in Figure\,\ref{c4phs}, along with five model fits.

   After a search throughout our model parameter space, we found 
that it is impossible to fit all features with a opaque-disk model {\bf and}
a small height $h$. The essence of the problem is that all such models 
produce an $``M"$-like feature centered at $\phi$ = 0.0 or 0.55 in 
the equivalent width-phase curve. This minimum occurs because the 
{\bf absorption} from the disk is greater just before/after the central 
conjunction. At the former times the 
disk appears canted to an external observer, and hence, like Saturn's 
rings often viewed from Earth, has a relatively large projected area. 
Conversely, for opaque disk models with large values of $h$
the absorption reaches a flat plateau well before central occultation.  
The absence of either a pointed or M-like absorption profile maximum in 
Fig.\,\ref{c4phs} drives our modeled disk semiheight $h$ to intermediate
values of 0.5--1R$_*$. Our best models suggested 
a clean break between the inner edge of the disk and the star.

  The disk parameters have different degrees of influence on various
portions of the phase curve. The shape of the peak ($M$-like, triangular,
or a flat plateau) is largely determined by the disk height.
The narrowness/broadness of the core component of the absorption profile
is determined by the arctangent of the disk's semiheight $h$ and 
its outer radius $r_o$. Thus, these two parameters are coupled.  A uniform
disk connected to the star, with $r_i$ $\approx$ 1R$_*$, would cause the
absorption to appear soon after the face-on phase, i.e., at $\phi$ $<$ 0.80,
The general absorption profile would then acquire a triangular shape that
is not observed. However, none of these parameters in the opaque model 
representation can explain the ingress tail in Fig.\,\ref{c4phs}. 
The failure to match the tail worsens if one departs 
from the simple $i$ = 90$^o$, $\beta$  = 90$^o$ geometry. 
{\bf Reducing the magnetic angle $\beta$ does not change the shape of the
curve but deemphasizes the effect of a given separation of the disk from the 
star, typically by $\sim$30\%. Observing the star from a smaller $i$ aspect
does even more of the same, but more importantly it triangularizes the profile 
such that there is no turn over or plateau near the maximum, or tail extension
at the minimum (limb contact).} Likewise, the neglect of limb darkening
deemphasizes any absorption tail. To remedy the deficiencies
of the single-component opaque model in Fig.\,\ref{c4phs}, we
developed the two-component disk model described above. 

  As the disk rotates from its face-on orientation in the two-component 
model, absorption from the disk first is evident as
the peripheral translucent component advances over the star's limb, in 
advance of the opaque edge. The difference in the absorption curves between 
the one and two-component models is substantial. In Figure\,\ref{c4phs}
we show best fit models for $r_o$ = 5R$_*$, 10R$_*$, and 40R$_*.$ 
As with the one-component models, we found that fitting the data with 
large $r_o$ requires that the $r_i$ values 
be made small until finally the inner edge of the disk reaches the star's 
surface. However, for extensive disks, e.g. $r_{o}$ = 40R$_{*}$, the
semiheight of the disk would be nearly 15 R$_{*}$ if the disk reached
down to the star.  We can be skeptical that this class of large-disk models 
is realistic for real Bp disks.  A second practical consideration is that 
models producing tolerable fits must have not only large $h$ values but 
proportionately even larger heights $h_{S}.$ Just as with the former class of
opaque models, one must invoke separate arguments to discard the possible 
solutions with largest and smallest disk radii. To address whether disks
can be much thicker or thinner in $h$ than 1R$_*$ will require
sophisticated hydrodynamic models that include radiative cooling and
the effects of post-shock flows.

  Interestingly, our two component models require that the ratio $h_S/h$ be
close to 2.5--3, i.e., the value of the exponential constant $e$ itself. 
This fact tends to validate the basic picture that the C\,IV absorptions
are optically thick and become optically thin at 2.5--3 scale heights.
The implied value of $\tau_{\lambda1548}$ $\sim$ 10 in the disk plane 
is also within only a small factor (perhaps 20\%) of the value obtained in 
the analysis of the C\,IV absorptions represented in Table\,2. 
This agreement suggests that
our crude opaque-translucent models give a reasonably accurate 
representation of the details of the absorption profile in Fig.\,\ref{c4phs}.

   The two component models gave us a freedom to explore a range of $h$ 
values much larger than 1R$_*$. This became apparent in our trials for models 
similar to those shown in Fig.\,\ref{c4phs}, with $r_o$ $\ge$ 10R$_*$. As
shown in the figure, $h$ values as small as ${\frac 14}$R$_*$ blur the 
$``M$ minimum" at $\phi$ $\approx$ 0.00 to invisibility.
We were able to fit the data with (opaque component) $h$ 
values in the broad range of ${\frac 14}$\,--1R$_{*}.$

%FIGURE 7 old
   \begin{figure*}[t]
   \centering
   \includegraphics[width=11.5cm,angle=90]{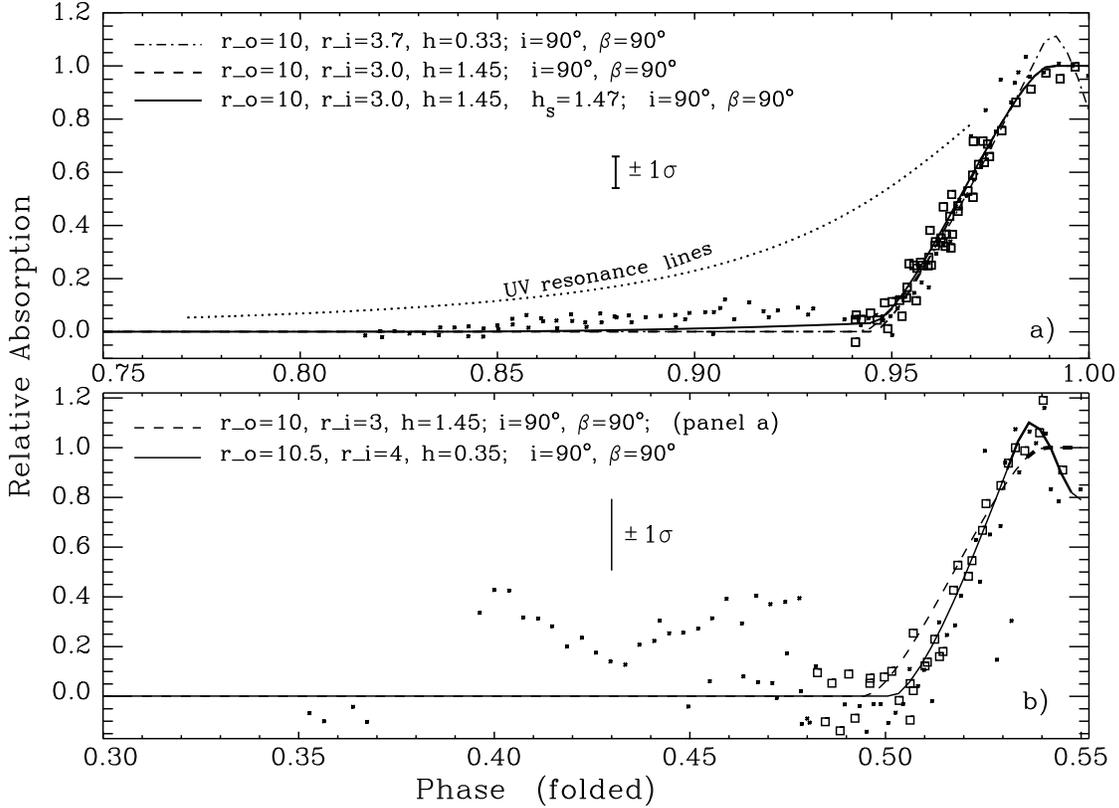}
\caption{
{\it Panel a:~} 
Fits to excess H$\alpha$ absorption with phase for the primary
occultation of 36\,Lyn centered at $\phi$ = 1.0, similar to
Fig.\,\ref{c4phs} for C\,IV.
Coded lines are fits for indicated disk dimensions computed from 
one-component (dashed) and two-component (solid) disk-absorption models.
Dots and squares represent observations made before and after mid-occultation,
respectively. Mean {\bf 1$\sigma$ error bars refer to the 2$\sigma$ taken 
from Paper 1. }
{\it Panel b:~} Fits to H$\alpha$ data responsible for the secondary 
occultation centered near $\phi$ = 0.55. The best fit model from panel {\it a},
with semiheight $h$ = 1.45R$_{*}$, is shown to be too thick for this disk 
sector.  The scaling is normalized to unity for the maximum absorption and
thus is enlarged relative to the scale of {\it panel a.}
}
\label{haphs}
\end{figure*}

\subsection{Fitting of H$\alpha$ variations with phase}
\label{alffit}

  The $T_{disk}$ = 7,500\,K and column density for the cool disk component 
determined from the iron lines is sufficient to make the optical 
depth in H$\alpha$ $\sim$5, thereby satisfying one of the assumptions
of our occultation models.

  In contrast to the UV resonance lines, it is relatively 
easy to fit H$\alpha$ absorption data even with our opaque 
component disk. In Figure\,\ref{haphs} we exhibit the H$\alpha$ 
absorption profiles for the two occultations of the rotation cycle. 
To match the zero point of the equivalent width data to the models,
we subtracted a plateau level of the equivalent width average,
1.34\,\AA~ at nonoccultation phases and scaled the peak absorption value
to 1.0 (all flux removed by the disk) for 
comparison with the models.  Because the data on either side of the maxima
seems to follow a common relation, we reflected the data around the
centroid phase ($\phi$ = 0.00 and 0.55 phases).  

  Before proceeding to the modeling analysis, we digress to discuss 
two important attributes of the H$\alpha$ curves that demonstrate that
the assumption of {\it planarity} of our geometrical disk models cannot 
be quite right:

\begin{itemize}
\item 
%Near the symmetry axis the data in 
Fig.\,\ref{haphs}a exhibits a 
plateau near $\phi$ = 1.00. This plateau is more apparent for the dots, 
which are flat for phases 0.98--1.00. The squares {\bf (post-occultation
data)} beyond 0.01 fall off more rapidly from the peak at 0.0 
than the dots {\bf (pre-occultation data).} This fact hints at 
a difference in $h$ on either side of the central plane and may manifest 
a slight warping of the disk. 
\\

\item A faint but distinct absorption may be present at phases separated from
the central lobe maxima; see the feature in the phase range
 $\phi$ = 0.86\,--\,0.93.  A similar
out-of-plane absorption is evident at Fig.\,\ref{haphs}b at $\phi$ = 
0.40\,--\,0.48. These features can be interpreted as matter {\bf separated from}
the disk torus.  As an alternative to the planar disk geometry, Michel \& 
Sturrock (1974), Nakajima (1985), Preuss et al. (2004), and TO05 have suggested 
that the wind particles will collect in local equipotential regions. {\bf The
locus of this surface depends upon} the balance of magnetic, radiative, and 
may have relevance to these features.

\end{itemize}

As in the C\,IV analysis, the $h$ value drives the shape
of the central core absorption near $\phi$ = 1.00. Likewise, the
half-width and the slopes of the profile are sensitive
to the r$_i$ and r$_o$ 
radii, respectively. For small disk models, we
could achieve matches only with thin rings, that is $r_o$ $\approx$ $r_i$. 
However, for values of $r_{o}$ below 8 R$_{*}$, we had difficulty fitting
even thin rings to the data. For models with  $r_{o}$
$\ge$10R$_{*}$ the fitting was easy, and the ratio of inner to outer
radii quickly settled to about 30\% in Fig.\,\ref{haphs}a. 
The {\bf dot-}dashed line 
in this panel (just visible {\bf between} 0.90$\le\phi\le$0.95) gives a 
best fit for the opaque-component (only) family with $r_{o}$ = 10R$_{*}$.
For larger $r_{o}$ values, $r_{i}$ approximately scales with the same 
semiheight. Our geometric models by themselves do not allow us to rule
out larger disks.

  To examine the influence of a thin (in direction perpendicular to the plane)
disk, we have also computed a sample ``M-shaped" core in Fig.\,\ref{haphs}a 
for a model with a semiheight $h$ = ${\frac 13}$R$_{*}$. This model is 
represented by the dot-dashed line. This attempt clearly does not fit the 
data for the primary occultation. Indeed, the best semiheight we found was 
$h$ = 1.45R$_{*}$, which is probably within the errors of the $h$ value
determined from C\,IV absorptions.
The secondary occultation event represented in Fig.\,\ref{haphs}b 
tells a different story. First, the absolute maximum of this absorption
curve is only 33\% as strong as the absorption in the primary occultation.
Consequently, the effects of noise are much larger.
Nonetheless, inspection shows that this maximum has almost
exactly the same half-width and fall-off gradient with phase as the
primary. One difference with respect to the primary
maximum is that the H$\alpha$ absorption curve exhibits
a possible M-shape, reaching down some 25\% from the maximum height 
at $\phi$ $\approx$ {\bf 0.538.} The significance of
this inference is drawn from five observations in this diagram, of which four 
are much lower than unity by more than {\bf the} error bars. 
In Fig.\,\ref{haphs}b we show both our best fit for the primary maximum 
occultation in Fig.\,\ref{haphs}a (dashed line) and the best fit for the 
secondary maximum, this time with a semiheight $h$ = 0.35R$_*$.
The best $r_o$ and $r_i$ values for both occultations are practically the same.
According to this result, the two opposite sectors are similar, with the 
important difference that the sector responsible for the secondary 
occultation is four times thinner in height.

  How does our finding that H$\alpha$ is formed closer to the
plane than C\,IV fit into our current understanding of the disk 
dynamics?  In the Donati et al. (2002) and uO02 models, newly shocked wind 
particles cool and settle into a thin disk. The H$\alpha$-absorbing volume
can be expected to overlap the cool component we identified from our Fe 
line analysis.  Our standard model of the cool component
predicts an excess H$\alpha$ absorption, for full-disk occultation, of 
0.18\AA.~  This value happens to be the same as the peak excess absorption 
given in Paper 1. This agreement is fortuitous, considering that 
we have compared moderately optically thick H$\alpha$ and optically thin 
Fe\,II lines in (probably) only partially overlapping volumes.

\section{Conclusions}

  We can recap the
%significant 
conclusions of this paper as follows:

\begin{enumerate}

\item Like the circumstellar disks of early-type Bp stars, the 
   static ``disk" of 36\,Lyn consists of both a cool and a hot component.
   In general terms, this confirms the hydrodynamic result of Babel \& 
   Montmerle (1997) for IQ\,Aur. The cool component of 36\,Lyn is observed 
   in the Fe-group lines (mainly Fe$^{2+}$)
   arising from low excitation states; the hot component is visible in
   the UV resonance lines. The volume in which H$\alpha$ absorption is 
   formed probably overlaps both these regions, but we do not know the 
   extent of this overlap.

\item We have determined the temperature of the cool disk component to be
   7,500$\pm{500}$\,K from an analysis of Fe\,II and Fe\,III lines variations
   during primary disk occultation.
   The temperature ratio $T_{disk}$/$T_{eff}$ $\approx$0.6 for 36\,Lyn. 
   This is {\bf also} the ratio SG01 found for the hot Bp stars $\sigma$\,Ori\,E 
   and $\beta$\,Cep.

\item The physical parameters of this disk, including its column density,
   insure that the strongest low-excitation lines are readily visible
   in {\it IUE} spectra, but they are not quite sufficient to allow the
   disk component to be visible in even strong optical Fe lines.

\item The ionization temperature(s) determined from the presence of
   disk N\,V and C\,IV absorptions reveal that part of the disk is
   heated, perhaps to 30,000\,K or more. For C\,IV there is only a hint 
   of emission at non-occultation phases, unlike the disks of hotter
   Bp stars. 

\item The relative heights of the absorption maxima corresponding
   to the occultations by two disk segments differ among the highly
   excited ``super ions" (N\,V, C\,IV, and Si\,IV) themselves and with 
   respect to resonance lines of less excited ions. The two opposing 
   disk segments do not exhibit the same ranges of ionization.  The 
   disk sector responsible for absorptions at $\phi$ = 0.00 has a more 
   restricted temperature range than the sector visible at $\phi$ 
   $\approx$ 0.55. The existence of restricted temperatures 
   implies that the production of the ``super ions" does 
   not result from the irradiation of gas by local soft X-rays.

\item Geometric modeling of the occultation from the UV resonance and 
   H$\alpha$ lines suggests that the inner edge of the disk (ring) 
   is detached from the star's surface. The inner edge is is located at
   $\approx$ ${\frac 13}$$r_o$.  This conclusion 
   is secure for H$\alpha$ but is less secure for the UV data. 

\item Analysis of the equivalent width maxima of the $H\alpha$ shows
   clearly that the associated opposing disk sectors have
   nearly the same inner and outer disk values. They differ by a factor 
   of 3 in their heights above the magnetic plane. The secondary absorption 
   appears to be weaker because the disk sector responsible for it is
   thinner (measured out of the disk plane) than the opposing segment by 
   about the same factor.

\item Like H$\alpha$, the UV resonance lines are opaque near the central 
   plane. Their absorptions are formed in volumes having nearly the same 
   range of radii as the H$\alpha$ absorption (i.e., from 3R$_*$ to 10R$_*$). 
   The UV resonance lines indicate significant additional 
   translucent extensions beyond this {\bf height,} 
   with  h$_{S}$/$h$ $\approx$ 3. 

\end{enumerate}

\acknowledgements{This work was supported in part by NASA grant NNG04GE75E.
The optical data were obtained using the MuSiCoS spectropolarimeter at
Pic Du Midi Observatory, the Canada-France-Hawaii Telescope, operated by 
the National Rsearch Council of Canada, the Centre National de la Recherche
Scientifique of France, the University of Hawaii, the Dominion 
Astrophysical Observatory, Herzberg Institute of Astrophysics, and the
David Dunlap Observatory, University of Toronto. 
C. T. Bolton's research was partially supported
by a Discovery Grant from the Natural Sciences and Engineering
Research Council of Canada.  John Liska and Mel Blake assisted
in obtaining and reducing the DDO spectra. 
}

\end{document}